\documentclass[pre,aps,preprint,showpacs,amsmath,amssymb, superscriptaddress]{revtex4-1}

\usepackage{graphicx}
\begin{document}

\title{Shape dependent finite-size effect of critical two-dimensional Ising model on a triangular lattice}
\author{Xintian Wu}
\email{wuxt@bnu.edu.cn}
\affiliation{Department of Physics, Beijing Normal University,
Beijing, 100875, China}

\author{Nickolay Izmailian}
\email{izmail@yerphi.am}
\affiliation{A.I. Alikhanyan National
Science Laboratory, Alikhanian Br.2, 375036 Yerevan, Armenia.  }
\author{ Wenan  Guo }
\email{waguo@bnu.edu.cn}
\affiliation{Department of Physics, Beijing Normal University,
Beijing, 100875, China}

\date{\today}

\begin{abstract}

Using the bond-propagation algorithm, we study the finite-size behavior of the
critical two-dimensional Ising model on a finite triangular lattice with free
boundaries in five shapes: triangle, rhombus, trapezoid, hexagon and
rectangle. 
The critical free energy, internal energy and specific heat are calculated.
The accuracy of the free energy reaches $10^{-26}$.
Based on accurate data on several finite systems with linear size up to
$N=2000$, we extract the bulk, surface and corner parts of the free energy,
internal energy and specific heat accurately. We confirm the conformal field
theory prediction of the corner free energy to be universal and
find logarithmic corrections in higher order terms
in the critical free energy for the rhombus, trapezoid, and
hexagon shaped systems, which are absent for the triangle and rectangle
shaped systems.
The logarithmic edge corrections due to edges parallel or perpendicular to
the bond directions in the internal energy are found to be identical,
while the logarithmic edge corrections due to corresponding edges
in the free energy and the specific heat are different.
The corner internal energy and corner specific heat for angles
$\pi/3$, $\pi/2$ and $2\pi/3$ are obtained, as well as higher order
corrections.
Comparing with the corner internal energy and corner specific
heat previously found on a rectangle of the square lattice (Phys. Rev. E. {\bf 86}
041149 (2012)), we conclude that the corner internal energy and corner
specific heat for the rectangle shape are not universal.
\end{abstract}

\pacs{02.70.-c, 05.50.+q, 75.10.Nr, 75.10.Hk}

\maketitle

\section {Introduction}
The finite-size scaling theory, introduced by Fisher, finds extensive
applications in the
analysis of experimental, Monte Carlo, and transfer-matrix data,
as well as in recent theoretical developments related to conformal
invariance \cite{privman,privman1,cardy,blote}. It becomes of
practical interest due to the recent progresses in fine processing
technologies, which has enabled the fabrication of nanoscale
materials with novel shapes \cite{kawata,puntes,yin}.
Exact solutions have been playing a key role in determining the
form of finite-size scaling. Ferdinand and Fisher
\cite{fisher1969} pioneered on the two-dimensional (2D) Ising
model on a finite size square  lattice, which extended Onsager's
exact solution \cite{onsager} and stimulated the ideas of
finite-size scaling. Since then, exact results of the model on
finite size lattices with various boundaries have been studied
intensively
\cite{onsager,kaufman,fisher1969,fisher,izmailian2002a,izmailian2002b,izmailian2007,salas,izmailian1,Janke,newell}.
Detailed knowledge has been obtained for the torus case
\cite{izmailian2002a}, for the helical boundary condition
\cite{izmailian2007}, for the Brascamp-Kunz boundary condition
\cite{izmailian2002b,Janke} and for an infinitely long cylinder
\cite{izmailian1}. The exact solution on the triangular lattice has
also been studied intensively \cite{newell,salas}.

However for the 2D Ising model the exact solution with free boundaries,
i.e. free edges and sharp corners, is still missed. As we know,
the Bethe ansatz is a powerful technique to solve the 2D Ising
model, but this technique would not allow to place the system on a
rectangle with free boundary conditions on top and bottom, since
the expression of the boundary state in terms of the Bethe
eigenvectors is unknown, which is a famous unsolved problem in
statistical mechanics.  Although there are Monte Carlo and
transfer matrix studies on this problem \cite{landau, stosic}, the
accuracy or the system sizes of the results are not enough for
extracting the finite size corrections. Meanwhile, for 2D critical
systems, a huge amount of knowledge has been obtained by the
application of the powerful techniques of integrability and
conformal field theory (CFT) \cite{blote, cardy, kleban,jacobsen1,
jacobsen2}. Cardy and Peschel predicted that the next subdominant
contribution to the free energy on a square comes from the
corners \cite{cardy}, which is universal, and related to the central
charge $c$ in the continuum limit. Kleban and Vassileva
\cite{kleban} extended the result to a rectangle. These results
are consistent with the conjectured exact analytic formula for the
Ising model on a square lattice \cite{jacobsen3}.

Several years ago  an efficient bond propagation (BP) algorithm
was developed for computing the partition function of the Ising
model with free edges and corners in two dimensions
\cite{loh1,loh2}.  Making use of this algorithm, we recently 
determined numerically the exact partition function of the Ising model on the
square lattice with rectangle shape and free boundaries \cite{wu}.
We not only confirmed the CFT predictions, but also found
logarithmic corrections due to corners in the internal energy and specific heat.

\begin{figure}
\includegraphics[width=0.5\textwidth]{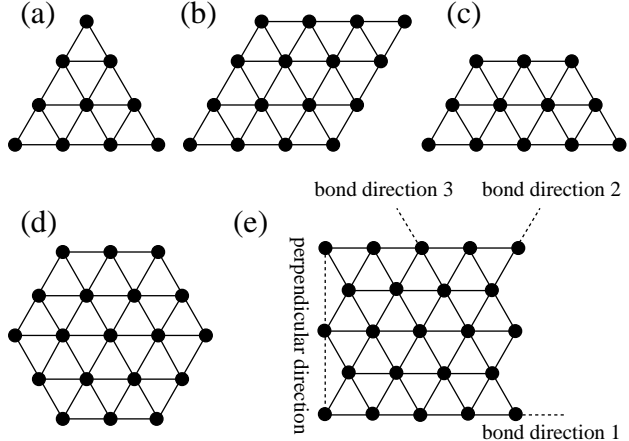}
\caption{ (a) The triangle-shaped triangular lattice with $N=4$. (b) The
rhombus-shaped lattice with $N=4$.    (c) The trapezoid-shaped
lattice with $N=3$.
(d) The hexagon-shaped lattice with $N=3$. (e) The rectangle-shaped lattice with $N=5$.
Three bond directions and the perpendicular direction are shown (see text).}%}
\label{shape}
\end{figure}

In present paper we apply the BP algorithm to study the Ising model
on finite triangular lattices with free boundaries in five different shapes,
focusing on how the shape affects the finite-size scaling.
The five shapes are triangle, rhombus, trapezoid, hexagon and rectangle, as 
shown in Fig. \ref{shape}.
Based on accurate data on a sequence of finite systems with linear size up to
$N=2000$, we extract the bulk, surface and corner parts of the free energy,
internal energy and specific heat accurately.
We verify the conformal field theory prediction of the corner free energy and
find logarithmic corrections in higher order terms in the critical free energy 
for the rhombus, trapezoid, and hexagon shaped systems, which are absent for 
the triangle and rectangle shaped systems.
The logarithmic edge corrections due to edges parallel or perpendicular to
the bond direction in the internal energy are found to be identical,
while the logarithmic edge corrections due to the corresponding edges
in the specific heat are different.
The corner internal energy and corner specific heat for angles
$\pi/3$, $\pi/2$ and $2\pi/3$ are obtained, as well as higher order
corrections.
Comparing with the previous found corner internal energy and corner specific
heat on a rectangle of the square lattice \cite{wu}, we conclude that the corner 
internal energy and corner specific heat for the rectangle shape are not universal.

Our paper is organized as follows: In section \ref{method}, we briefly
describe the BP algorithm used here.  We present in sec. \ref{results}
our main results and discussion. We conclude in sec. \ref{conclusion}.

\section{Method}
\label{method}

The partition function of the Ising model on the 2D triangular lattice is
\begin{equation}
Z=\sum_{\{\sigma_i\}}\exp{(\beta \sum_{\langle i,j\rangle }\sigma_i\sigma_j)}
\label{eq:partition},
\end{equation}
where the nearest neighbor couplings are dimensionless and $\beta$
is the inverse temperature.
This partition function for finite triangular lattice with five
different shapes and open boundaries is calculated
with the BP algorithm \cite{loh1} at the exact critical point
$\beta_c=\frac{1}{4}\ln(3)=0.274653072167\cdots$.
Figure \ref{shape} shows the five shapes: triangle, rhombus, trapezoid,
hexagon and rectangle.
The linear size  $N$ of a finite lattice is
defined as the length of edges in the triangle, rhombus and hexagon
cases, of which the length of edges are equal. For the trapezoid shape, $N$ 
is the shortest edge, of which the lengths of the three
short edges are required to be equal. For the rectangle, $N$ is
defined as the length of the bottom edge, and the number of layers is
required to be $N$,  which means the actual geometrical
vertical length is $N\sqrt{3}/2$. According to the finite-size
scaling \cite{privman}, the system size should be the actual
geometrical length, not the number of layers. Therefore the aspect
ratio of the rectangle, we consider here, is $\sqrt{3}/2$ rather
than $1$.

\begin{figure}
\includegraphics[width=0.5\textwidth]{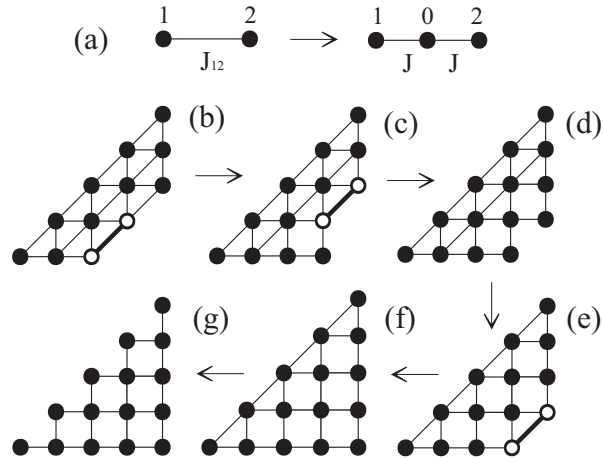}
\caption{ (a) Inverse of the BP series reduction. See the formulation
for this transformation in text. (b)-(g) are schematic of the
algorithm for the trapezoid-shaped lattice. From (b) to (c), the
inverse of BP series reduction is applied to the diagonal bond (
thick line) at the bottom of the trapezoid. From (c) to (d), this
operation is applied to another diagonal bond (thick line). From
(d) to (e) bond propagation operations are applied. From (e) to
(f), the inverse of BP series reduction is applied again. (f) to
(g) Those diagonal bonds are eliminated by the usual BP procedure.
} \label{bp}
\end{figure}

The BP algorithm for the Ising model on the triangular lattice has been
described in detail in \cite{loh2}. For the triangle-shaped and
rhombus-shaped lattice, this algorithm can be applied directly. However, for
the trapezoid-shaped, hexagon-shaped and rectangle-shaped lattice, the
inverse of the BP series reduction should be introduced, 
which corresponds to generating a
new spin between two spins such that
\begin{equation}
e^{J_{12}\sigma_1\sigma_2}\equiv \sum_{\sigma_0}e^{\delta
F+J\sigma_0(\sigma_1+\sigma_2)} , \label{eq:series}
\end{equation}
as illustrated in Fig. \ref{bp}(a).
It is convenient to use variables
$j \equiv e^{-J},j_{12} \equiv e^{-J_{12}}$ and $\delta f \equiv e^{\delta F}$. Then we
get the solution $\delta f= j_{12}/2$ and
$j=\sqrt{\frac{1}{j^2_{12}}-\sqrt{\frac{1}{j_{12}^{4}}-1}} $.

In each step of the BP algorithm the transformation is exact. The
numerical accuracy is only limited by the machine's precision, which is the
round-off error $10^{-32}$ in the quadruple precision. The BP
algorithm needs about $N^3$ steps to calculate the free energy of
a $N\times N$ lattice (much faster than other numerical method).
Therefore the total error is approximately $N^{3/2}\times
10^{-32}$. This estimation has been verified in the following way:
We compared the results obtained using double precision, in which
there are 16 effective decimal digits, and those using quadruple
precision. Because the latter results are much more accurate than
the formal, we can estimate the error in double precision results
by taking the quadruple results as the exact results. We thus
found that the error is about $N^{3/2} \times 10^{-16}$. In our
calculation, the largest size reached is $N=2000$, the round-off
error is less than $10^{-26}$.

The free energy density, internal energy per spin and specific
heat density are calculated at the critical point according
to
\begin{equation}
f=\frac{\ln Z}{S}, \hskip 0.5cm u=\frac{\partial f}{\partial
\beta}, \hskip 0.5cm c=\beta^2 \frac{\partial^2 f}{\partial
\beta^2},
\end{equation}
respectively, where $S$ is the number of spins on the lattice,
which is $S=N(N+1)/2$, $N^2$, $N(3N-1)/2$, $3N^2-3N+1$, $N^2-(N-1)/2$ for the
triangle, rhombus, trapezoid, hexagon and rectangle shaped system,
respectively.
One can alternatively define the free energy density according to
the actual geometrical area, which is different from the present 
definition by a trivial constant.

With the BP algorithm, we obtain the free energy density $f$ directly. % Then we
The internal energy and specific heat are calculated by using a
differentiation method
\begin{eqnarray}
u & \approx & -\frac{f(\beta_c+\Delta \beta)-f(\beta_c-\Delta
\beta)}{2\Delta \beta} ,\nonumber \\
 c & \approx & \beta_c^2\frac{f(\beta_c+\Delta \beta)+f(\beta_c-\Delta
\beta)-2f(\beta_c)}{(\Delta \beta)^2}.
\end{eqnarray}
In our calculation, $\Delta \beta=10^{-7}$ is used.
The analysis of error in the calculations of  $u$ and $c$ has been discussed in
\cite{wu}.
The final estimation of the accuracy of the free energy,
internal energy and specific heat is $10^{-26},10^{-11},10^{-9}$,
respectively.

The calculations were carried out for $105, 104, 103, 85, 127$
systems, with linear size $N$ varying from $30$ to $2000$, for the
triangle, rhombus, trapezoid, hexagon and rectangle shaped
triangular lattice, respectively.

\section{Results}
\label{results}
\subsection{Critical free energy density}
By fitting the finite size data,
we find that the exact expansion of the critical free energy can
be written in the following form, with $k$ from 3 to 12 for the 
triangle and rectangle, $k$ from 3 to 9 for the other shapes:
\begin{equation}
f=f_{\infty}+f_{surf}\frac{p(N)}{S}+\frac{f_{corn}\ln
N+f_2}{S}+\sum_{k=3}\frac{f_{k}+l_k \ln N}{S^{k/2}},
\label{freeenergy}
\end{equation}
where $p(N)$ is the perimeter, which equals to $3N, 4N, 5N-1, 6N, (2+\sqrt{3})N$ for the
triangle, rhombus, trapezoid, hexagon and rectangle, respectively.
This expansion is different from that for the triangular lattice with periodic
boundary conditions \cite{salas}, in which there's no surface, corner terms,
logarithmic corrections, and only even $k$ presents.

The fitting method is the standard Levenberg-Marquardt method for nonlinear
fit. The standard deviation (SD) is defined by
$\sigma=\sqrt{\sum_i(f_i-f_i^{(fit)})^2/(n_d-n_f)}$ with $f_i$ the
numerical data, $f_i^{(fit)}$  the value given by the fitting
formula, $n_d$  the number of data used and $n_f$ the number of fitting
parameters. For all cases, $\sigma$ reaches $10^{-25}$. The accuracy is seen
from the fitted bulk free energy density $f_{\infty}$: the worst
fit among the five cases is for the rhombus, which yields
$f_{\infty}=0.8795853861615715170938962(7)$, and the
best one is for the triangle yielding
$ f_{\infty}=0.87958538616157151709389605(3)$. These results  coincide with
the exact value $f_{\infty}=0.8795853861615715170938960283\cdots$
 \cite{newell}
in more than 24 decimal numbers.

\begin{table*}[htbp]
 \caption{ The fitted edge, corner free energy and $f_2$ in Eq. (\ref{freeenergy}).  }
\begin{tabular}{llll}

\hline
shape     & $f_{surf}$ & $f_{corn}$ & $f_2$  \\
\hline

triangle  & $-0.1030776388340906553343(2)$ & $0.1666666666666666667(4)$ & $0.006804832446685952(4)$    \\
rhombus   & $-0.103077638834090655334(2)$  & $0.14583333333333333(2)$   & $0.1839728334687587(1)$   \\
trapezoid & $ -0.103077638834090655335(1)$ & $0.14583333333333335(2)$   & $0.2213952601428039(1)$    \\
hexagon   & $-0.103077638834090655335(2)$  & $0.10416666666666668(3)$   & $0.4758271527774812(2)$ \\
rectangle & $-0.109078407337305392239(2)$  & $0.125000000000000001(2)$  & $0.22549835683994714(1)$   \\

\hline
\end{tabular}
\label{fitfen1}
\end{table*}

According to finite-size scaling, the surface correction term $f_{surf}p(N)/S$
stems from free edges. As shown in Fig. \ref{shape}(e), there are three bond
directions for the triangular lattice. For the triangle, rhombus,
trapezoid, and hexagon shapes, the edges are all along one bond
direction. The edge free energy per unit length on these edges
should be equal. For the rectangle shape, there are two edges
along one bond direction and two other edges perpendicular to that
bond direction in a zigzag way. Therefore the edge free energy per unit length
along the two different directions can be different in principle.
To be clear, 
we denote the surface free energy per unit length (the straight length,
not the total length of the zigzag line)  along and
perpendicular to the bond direction as $f_{surf}^{\parallel}$ and
$f_{surf}^{\perp}$, respectively. The surface free energy is thus
$f_{surf}p(N)= f_{surf}^{\parallel} p(N)$
for the triangle, rhombus, trapezoid, and hexagon shape, respectively. For
the rectangle shape, it is $f_{surf} P(N)
=\bar{f}_{surf} (2+\sqrt{3})N$, with
$\bar{f}_{surf}=(2 f_{surf}^{\parallel}+\sqrt{3} f_{surf}^{\perp})/(2+\sqrt{3})$, which
is the mean surface free energy per unit perimeter,
considering the straight length of the edges along the
perpendicular direction is $\sqrt{3}N/2$, which is used to define the perimeter. 
The fitted values of
$f_{surf}$ is given in Tab. \ref{fitfen1}. For the rectangle
case, we obtain
\begin{equation}
f_{surf}^{\parallel}=-0.103077638834090655334(2), \hskip 1cm
    f_{surf}^{\perp}=-0.116007497958656704304(2).
\end{equation}

Cardy and Peschel showed in \cite{cardy} that the presence of a
corner of interior angle $\gamma$ along boundaries of typical size
$N$ give rise to a logarithmic correction to the critical free
energy density
\begin{equation}
f_{corn}^{(\gamma)}\frac{\ln
N}{S}=-\frac{c\gamma}{24\pi}(1-(\frac{\pi}{\gamma})^2)\frac{\ln
N}{S}, \label{gamma}
\end{equation}
where $c$ is the conformal anomaly.  The total corner free energy is

\begin{eqnarray}
f_{corn}=\left\{\begin{array}{lcl} 3 f_{corn}^{(\pi/3)}=\frac{1}{6}=0.16666666... \hspace{3.6cm} \mbox{for triangle} 
\nonumber\\
2\left(f_{corn}^{(\pi/3)}+f_{corn}^{(2\pi/3)}\right)=\frac{7}{48}=0.145833333... 
\hspace{1cm} \mbox{for rhombus and trapezoid} \nonumber\\
6 f_{corn}^{(2\pi/3)}=\frac{5}{48}=0.104166666... \hspace{3.1cm} \mbox{for hexagon} 
\nonumber\\
 4 f_{corn}^{(\pi/2)}=\frac{1}{8}=0.125 \hspace{5cm} \mbox{for rectangle} \nonumber
\end{array}
\right.
\end{eqnarray}
The fitted corner free energy $f_{corn}$ for the
five shapes are listed in Tab. \ref{fitfen1}, from which one can see
that our results reproduces the CFT result very accurately.

\begin{table*}[htbp]
\caption{ The fitted parameters for order $k \ge 3$ in Eq.  (\ref{freeenergy})
for the critical free energy
of the triangle and rectangle shaped triangular lattices. There is no logarithmic corrections
except for the corner term, i.e., all $l_k=0$. }
\begin{tabular}{lll}

\hline
 shape   & triangle & rectangle \\
\hline

$f_3 $    & $0.117851130197757931(2)$ & $-0.0144959340650645(9)$   \\
$f_4 $    & $0$                       & $ 0.0180683705511(1)$  \\
$f_5 $    & $-0.00245523187923(4)$    & $-0.01237101066(1)$ \\
$f_6 $    & $0.001247090597(1)$       & $ 0.0168641489(1)$ \\
$f_7 $    & $-0.0016255469(3)$        & $ 0.00302743(4)$ \\
$f_8 $    & $0.00124715(1)$           & $ 0.001335(1)$  \\
$f_9 $    & $-0.0005633(2)$           & $-0.03133(3)$  \\
$f_{10} $ & $-0.000553(3)$            & $ 0.0701(4)$\\
$f_{11} $ & $0.00159(2)$              & $-0.161(2)$\\
$f_{12} $ & $-0.050(4)$               & $ $\\
\hline
\end{tabular}
\label{fitfen2}
\end{table*}

\begin{table*}[htbp]
 \caption{ The fitted parameters of Eq.  (\ref{freeenergy}) for the critical free energy. The third and fourth
 order logarithmic correction $l_3,l_4$ are zero. }
\begin{tabular}{llll}

\hline
 shape   & rhombus & trapezoid & hexagon \\
\hline

$f_3 $  & $0.04861111111112(2)$ & $-0.00210692865756(2)$  & $-0.06014065304055(7)$  \\
$f_4 $  & $-0.00810185186(2)$   & $ 0.01363281706(2)$     & $0.03472222217(7)$  \\
$f_5 $  & $0.00381144(6)$       & $-0.00686168(6)$        & $-0.0180976(4)$  \\
$f_6 $  & $-0.00215(1)$         & $-0.00894(1)0$          & $0.0150(1)$  \\
$f_7$  & $-0.0008(5)$           & $0.0316(6)$             & $ -0.018(7)$  \\
$f_8 $  & $-0.026(4)$           & $-0.113(5)$             & $-0.56(9)$  \\
$f_9 $  & $0.032(2)$            & $0.097(3)$              & $0.21(3)$  \\
\hline
$l_5 $  & $0.00346339(1)$       & $0.00494874(1)$         & $0.00999797(6)$  \\
$l_6 $  & $-0.003461(2)$        & $-0.008767(3)$          & $-0.00863(2)0$  \\
$l_7 $ & $0.0024(1)$            & $ 0.0117(2)$            & $0.011(2)0$  \\
$l_8 $  & $0.006(2)$            & $ 0.011(3)$             & $0.17(5)$  \\
$l_9 $  & $0.007(5)$            & $ 0.070(8)$             & $0.9(2)$  \\

\hline
\end{tabular}
\label{fitfen3}
\end{table*}

In the previous study on the finite square lattice in a rectangle
shape\cite{wu}, we estimated the corner free energy $f_{corn}=0.125\pm 2.0\time 10^{-10}$
for the rectangle shape with various aspect ratios. Here we find the same
result $f_{corn}=0.125\pm 2.0\times 10^{-18}$ on
the rectangle-shaped triangular lattice. This indicates that the corner term of the
free energy is independent of the microscopic properties of the lattice.
Therefore we proved the CFT prediction that the corner free energy is
universal \cite{cardy}.
However, our calculations are for the aspect ratio fixed as $\rho=2/\sqrt{3}$
at present work. Thus it is not enough to
further verify Kleban's CFT predictions on the effect of the aspect ratio of
the rectangle-shaped lattice \cite{kleban}.

For higher order terms, we give fitted coefficients in
Tab. \ref{fitfen2} and \ref{fitfen3}.
For the critical free energy of the triangle and rectangle shaped triangular lattices,
there is no logarithmic corrections except for the corner term, i.e., all $l_k=0$.
For the triangle shape, the coefficient $f_4$ % $1/N^4$
is determined to be zero since it is extremely small in our fits if this
term is present,  and the fitting result changes little if we discard this term.
In contrast to the triangle and rectangle shape,
for the other three shaped triangular lattices, we find higher order
($k \ge 5$) logarithmic corrections besides the corner term.
Although these logarithmic corrections are very weak,
with very small coefficients, our high accurate data indicate their existence.

\subsection{Critical internal energy density}
We fit the data on the critical
internal energy  with the following formula
\begin{equation}
u=u_{\infty}+u_{surf}\frac{p(N) \ln N}{S} +u_{corn} \frac{\ln N
}{S} +\sum_{k=1}^{\infty}\frac{u_k}{S^{k/2}},
 \label{internalenergy}
\end{equation}
where $p(N)$ is again the perimeter.  In our fits, $k$ is truncated to $4$.
Again, this formula is different from that for the triangular lattice with periodic
boundary conditions \cite{salas}, in which there's no logarithmic surface term, no corner term,
and only odd $k$ presents.
The bulk value $u_{\infty}$ is known to be $2$ \cite{newell}. Our fit of
$u_{\infty}$ is $2.0\pm1.0\times 10^{-10}$ for the five shapes.
The other fitted parameter are given in Tab. \ref{fitien}.

\begin{table*}[hbtp]
\caption{The fitted parameters of Eq. (\ref{internalenergy}) for the
critical internal energy per spin.}
\begin{tabular}{llllll}
\hline
 shape   & triangle & rhombus & trapezoid & hexagon & rectangle\\
\hline

$u_{surf} $ & $-0.55132891(2)$  & $-0.551328867(8)$ & $-0.55132888(1)$ & $-0.55132891(2)$ & $-0.5513290(7)$  \\
$u_{corn} $ & $-1.653995(3)$    & $-0.735055(7)$    & $ -0.73506(1)$   & $1.10268(2) $    & $0.147707(6)$ \\
$u_1 $      & $0.1358131(3)$    & $-0.7286080(2)$   & $-1.4553110(3)$  & $ -3.4498392(7)$ & $ -1.1246224(2)$   \\
$u_2 $      & $-1.42006(1)$     & $-1.09359(3)$     & $ -0.55400(4)$   & $1.71014(4)$     & $-0.95485(2)$ \\
$u_3 $      & $ -0.70291(8)$    & $-0.2154(3)$      & $0.0302(4)$      & $-0.372(1)$      & $0.1049(2)$  \\
$u_4 $      & $-0.0243(4) $     & $0.113(2)$        & $0.044(4)$       & $0.49(1)$        & $-0.030(2)$ \\

\hline
\end{tabular}

\label{fitien}
\end{table*}
The leading correction is due to the edges (or surface), which is unknown
to our knowledge. We denote the
surface internal energy per unit length of the edge along one
bond direction by $u^{\parallel}_{surf}$, and that perpendicular to
that direction by $u^{\perp}_{surf}$. For the triangle,
rhombus, trapezoid, and hexagon shapes, we have
$u_{surf}=u_{surf}^{\parallel}$. For the rectangle we have
$u_{surf}=(2 u_{surf}^{\parallel}+\sqrt{3}u^{\perp}_{surf})/(2+\sqrt{3})$,
and
\begin{equation}
u_{surf}^{\parallel}=-0.55132889(2), \hskip 1cm
    u_{surf}^{\perp}=-0.55132895(7).
\end{equation}
This is an interesting result because it means that the surface
internal energy per unit length is symmetric for an edge along or
perpendicular to an arbitrary bond direction. Comparing this result with $u_{surf}$
for the rectangle shaped square lattice in our previous work\cite{wu}, where
$u_{surf}=0.6366198(1)$, we find that the ratio 
is $1.1547005$, which is very close to
$2/\sqrt{3}=1.154700538\cdots$. Because the exact value of
$u_{surf}$ for the square lattice is $2/\pi$ \cite{fisher}, we
conjecture that the exact value of the surface internal energy
for the triangular lattice is given by
\begin{equation}
u_{surf}^{\parallel}=u_{surf}^{\perp} =-\sqrt{3}/\pi.
\end{equation}

Following the convention for the critical free energy, we write the
coefficient of $(\ln N)/S$ as $u_{corn}$. We denote the corner
correction by $u_{corn}^{(\gamma)}$, where $\gamma$ is the angle of
the corner. Under the assumption that $u_{corn}$ is the sum
of the corner's contributions, we have 
\begin{eqnarray}
u_{corn}=\left\{\begin{array}{lcl}
3 u_{corn}^{(\pi/3)} \hspace{3.2cm} \mbox{for triangle} 
\nonumber\\
2\left(u_{corn}^{(\pi/3)}+u_{corn}^{(2\pi/3)}\right) \hspace{1cm} \mbox{for rhombus and trapezoid} \nonumber\\
6 u_{corn}^{(2\pi/3)} \hspace{3.1cm} \mbox{for hexagon} 
\nonumber\\
 4 u_{corn}^{(\pi/2)} \hspace{3.2cm} \mbox{for rectangle} \nonumber
\end{array}
\right.
\end{eqnarray}

From the Tab.
\ref{fitien}, we obtain the corner term for the three angles
\begin{equation}
u_{corn}^{(\pi/3)}=-0.551332(1),\hskip 1cm
u_{corn}^{(\pi/2)}=0.036927(2),\hskip 1cm
u_{corn}^{(2\pi/3)}=0.183781(3) .
 \label{eq:u-corn}
\end{equation}

In the previous study on the square lattice \cite{wu}, we obtained
the corner term $u_{corn}=-0.4502 (1)$ for the rectangle with various aspect ratios. 
It is different from the present result $u_{corn}=0.147707(6)$ for the 
rectangle-shaped triangular lattice, which indicates that the corner term of the 
internal energy depends on the microscopic structure of the lattice, thus is not 
universal.

The other parameters $B_1, B_2, B_3, B_4$ are also estimated and listed in
Tab. \ref{fitien}.  We have tried other forms of formula to
fit the critical internal energy. The terms $\ln S/S^{3/2}, \ln
S/S^2$ are excluded considering the coefficients are extremely
small. Moreover the standard deviations of the fits with these
terms included are much larger than those without them.

\subsection{Critical specific heat}
The data of the critical specific
heat are fitted using the following formula 
\begin{equation}
c=A_0\ln N+c_0+c_{surf} \frac{p(N)\ln N}{S} +c_{corn}\frac{\ln
N}{S} +\sum_{k=1}\frac{c_k}{S^{k/2}}, \label{specificheat}
\end{equation}
where $p(N)$ is the perimeter. $k$ is from 1 to $4$ in our fits.
Compared with the expansion for the triangular lattice with periodic
boundary conditions \cite{salas}, there are additional logarithmic surface term and
corner term.

The leading term $A_0\ln N$ is known from the exact result
\cite{salas}, which reads
$A_0=\frac{3\sqrt{3}}{4\pi}(\ln3)^2\approx 0.4990693780\cdots$.
Our fit yields $A_0\approx 0.499069374(5)$. The other fitted
parameters are listed in Tab. \ref{fitsh}.

The leading correction $p(N)\ln N/S $ is caused by the edges.
We denote the surface specific heat per unit length of the edge
along one bond direction by $c^{\parallel}_{surf}$, and that
perpendicular to the direction by $c^{\perp}_{surf}$. For
the triangle, rhombus, trapezoid, and hexagon shape, we have
$c_{surf}=c_{surf}^{\parallel}$,  for the rectangle, we set
$c_{surf}=(2 c_{surf}^{\parallel}+\sqrt{3}c^{\perp}_{surf})/(2+\sqrt{3})$, 
and find
\begin{equation}
c_{surf}^{\parallel}=0.166354(2), \hskip 1cm
    c_{surf}^{\perp}=0.105462(2).
\end{equation}
Note that this term is absent in the torus case \cite{fisher1969}
and not mentioned in the long strip case \cite{fisher}, but exists
in the cylinder case with Brascamp-Kunz boundary conditions
\cite{Janke,izmailian2002b}.

\begin{table*}[hbtp]
\caption{The fitted parameters of Eq. (\ref{specificheat}) for the
critical specific heat of the five shapes.}

\begin{tabular}{llllll}
\hline
 shape       & triangle         & rhombus          & trapezoid         & hexagon          & rectangle\\
\hline

$c_{0}    $  & $-0.80424237(1)$ & $-0.60510331(1)$ & $-0.51995011(1)$  & $-0.2673395(1)$  & $-0.573388895(9)$\\
$c_{surf} $  & $0.1663558(3)$   & $ 0.1663531(5)$  & $0.1663520(5)$    & $0.166338(9)$    & $0.1380941(8)$ \\
$c_1 $       & $-0.187862(5)$   & $-0.036692(3)$   & $-0.10391(2)$     & $-0.2749(2)$     & $-0.08529(9)$  \\
$c_{corn} $  & $0.7484(1)$      & $0.3864(2) $     & $0.3857(3)$       & $-0.343(6)$      & $-0.1510(2)$\\
$c_2 $       & $0.6410(4)$      & $0.6443(8)$      & $ 0.615(1)$       & $0.58(2)$        & $ 0.6833(7)$ \\
$c_3 $       & $ 0.339(2)$      & $0.177(6)$       & $0.04(1)$         & $-0.09(3)$       & $-0.057(5)$ \\
$c_4 $       & $0.024(8) $      & $-0.12(4)$       & $-0.00(7)$        & $1.8(3)$         & $0.17(3)$ \\

\hline
\end{tabular}
\label{fitsh}
\end{table*}

Following the convention for the critical free energy, we write the
coefficient of $(\ln N)/S$ as $c_{corn}$. We denote the corner
correction by $c_{corn}^{(\gamma)}$ where $\gamma$ is the angle of
the corner. Again, under the assumption that the total correction is the
sum of the corners,  we have 
\begin{eqnarray}
c_{corn}=\left\{\begin{array}{lcl}
3 c_{corn}^{(\pi/3)} \hspace{3.3cm} \mbox{for triangle} 
\nonumber\\
2\left(c_{corn}^{(\pi/3)}+c_{corn}^{(2\pi/3)}\right) \hspace{1cm} \mbox{for rhombus and trapezoid} \nonumber\\
6 c_{corn}^{(2\pi/3)} \hspace{3.1cm} \mbox{for hexagon} 
\nonumber\\
 4 c_{corn}^{(\pi/2)}\hspace{3.3cm} \mbox{for rectangle} \nonumber
\end{array}
\right.
\end{eqnarray}
From Tab. \ref{fitien}, we obtain the corner contribution of the three angles,
\begin{equation}
c_{corn}^{(\pi/3)}=0.24948(3),\hskip 1cm
c_{corn}^{(\pi/2)}=-0.03775(5),\hskip 1cm
c_{corn}^{(2\pi/3)}=-0.057(1) .
 \label{eq:c-corn}
\end{equation}

In the previous study on the square lattice \cite{wu}, we obtained
the corner term $c_{corn}=0.368(1)$ for the rectangle with various aspect ratios. 
It is different from the present result $c_{corn}=-0.1510(2)$ for the 
rectangle-shaped triangular lattice, which indicates that the corner term of the 
specific heat depends on the microscopic structure of the lattice, thus is not 
universal.

We have also tried other forms of fitting formula to fit the critical
specific heat data. For example, we added the terms $(\ln S)^3/S,
(\ln S)^2 /S $ in the fitting formula and found that their
coefficients are extremely small.

\section{Conclusion}.
\label{conclusion}
Using the BP algorithm, we have studied the 2D critical Ising
model on a triangular lattice with free boundaries. For five shapes,
triangle, rhombus, trapezoid, hexagon and rectangle, the critical
free energy, internal energy and specific heat have been calculated. We
have proved the conformal field theory prediction of the corner
free energy and have shown that the corner free energy, which is
proportional to the central charge $c$,  is indeed universal.
For the edges parallel or perpendicular to the
bond direction, the logarithmic edge corrections in the internal energy
have been found to be almost identical, while these corrections in the free
energy and in the specific heat have been found to be different.
Comparing with
the previous result on the square lattice in the rectangle shape, we have 
found that the corner internal energy $u_{corn}$ and the corner specific heat
 $c_{corn}$ for the rectangle shape are not universal, i.e., the
coefficient in front of $\ln N/S$ in the expansion of the internal
energy and the specific heat are different for the square lattice and
the triangular lattice.

We have also found that there exist logarithmic corrections in
higher orders, say, there are terms $\ln N/S^{5/2},\ln
N/S^3,\cdots$ in the critical free energy for the rhombus,
trapezoid, and hexagon shapes. However these terms are absent for
the triangle and rectangle shapes. This should be an interesting
subject to be further investigated.

\acknowledgments
This work is supported by the National Science
Foundation of China (NSFC) under Grant No. 11175018.

\end{document}